\documentclass[sigconf, nonacm]{acmart}

\AtBeginDocument{%
  \providecommand\BibTeX{{%
    \normalfont B\kern-0.5em{\scshape i\kern-0.25em b}\kern-0.8em\TeX}}}

\usepackage{bm}

\usepackage{enumitem}
\usepackage{makecell}
\usepackage{footmisc}

\newcommand{\resq}[2]{%
	\begin{itemize}
		\item[\textbf{RQ#1}:] #2
	\end{itemize}%
}

\usepackage[english]{babel}
\addto\extrasenglish{%
}

\usepackage[linesnumbered,lined,boxed,commentsnumbered, ruled, vlined]{algorithm2e}

\usepackage{multirow}
\usepackage{forest}
\usepackage{pgfplots}
\usetikzlibrary{decorations.markings, calc, positioning, backgrounds}

\tikzstyle{fm-topdown} = [font=\footnotesize\sffamily, parent anchor = south, child anchor = north,]

\tikzstyle{feature} = [rectangle, draw = black, fill = white, text = black, outer sep = 0pt, text depth = 0, text height = 1.25ex, inner sep = 1ex, font=\footnotesize\sffamily,]

\tikzstyle{legendFeature} = [rectangle, draw = black, fill = white, text = black, outer sep = 0pt, text depth = 0, text height = 1ex, inner sep = 1ex, font=\scriptsize\sffamily,]

\tikzstyle{abstract} = [fill = lightgray!25]

\tikzstyle{plain} = [draw=black]
\tikzstyle{mandatory} = [plain, decoration={
	markings,
	mark={at position 1 with {
			\node[circle, minimum size=4pt, inner sep = 0pt, draw=black, fill=black]{};}
	}
}, postaction = decorate]

\tikzstyle{optional} = [plain, decoration={
	markings,
	mark={at position 1 with {
			\node[circle, minimum size=4pt, inner sep = 0pt, draw = black, fill=white] {};}
	}
}, postaction = decorate]

\tikzstyle{optionalGreen} = [plain, decoration={
	markings,
	mark={at position 1 with {
			\node[circle, minimum size=4pt, inner sep = 0pt, draw = green, fill=white] {};}
	}
}, postaction = decorate]

\newcommand{\drawXOR}[4][]{
	\draw[fill=white, draw = black, line width = 0.5pt] (#3.south) -- ($(#3.south)!3mm!(#4.north)$) to[bend left] ($(#3.south)!3mm!(#2.north)$)  -- cycle;

}

\usepackage{colortbl}

\makeatletter
\renewcommand\@makefntext[1]{\leftskip=1em\hskip-1em\@makefnmark#1}
\makeatother

\begin{document}
\title{A Fast Counting-Free Algorithm for Computing Atomic Sets in Feature Models}

\author{Tobias He\ss}
\orcid{0000-0001-9389-9278}
\affiliation{%
	\institution{University of Ulm}
	\country{Germany}}

\author{Aaron Molt}
\affiliation{%
	\institution{University of Ulm}
	\country{Germany}}

\renewcommand{\shortauthors}{Heß and Molt}


\begin{abstract}
In the context of product-line engineering and feature models, atomic sets are sets of features that must always be selected together in order for a configuration to be valid. For many analyses and applications, these features may be condensed into one feature, without affecting, for instance, satisfiability, model counting, sampling, or knowledge compilation. However, the performance of current approaches tends to be insufficient in practice. This is especially true but not limited to approaches based on model counting. In this work, we present a counting-free algorithm for computing atomic sets that only relies on SAT solving. Our evaluation shows that it scales with ease to hard real-world systems and even succeeds for a contemporary version of the Linux kernel.
\end{abstract}

\maketitle
\section{Introduction}
\label{sec:intro}
Configurable systems, and the feature models modeling them, continue to increase in size and complexity~\cite{KTS+:ESECFSE17,T:SPLC20,SHN+:EMSE23}. While improvements in hardware and solving technologies tend to cushion some of this rise~\cite{FHS:CP20,FHI+:AIJ21,FHH:JEA21,SHN+:EMSE23}, preprocessing inputs and intermediate forms becomes increasingly more important~\cite{DHK:SPLC24,HSS+:SPLC24,FBAH20,BJK21,KKS+:ASE22,KHS+:VaMoS24,LM:AAAI14,S:SPLC08}. However, depending on the analysis or application, not all preprocessing techniques are equally viable~\cite{KKS+:ASE22}. For instance, techniques used in SAT solving predominantly only guarantee equisatisfiability~\cite{BJK21,KKS+:ASE22},\footnote{meaning that the processed output is satisfiable when the input was and vice versa} while altering other properties such as the number of satisfying solutions~\cite{KKS+:ASE22}.

While SAT solving is an important utility in feature-model analysis~\cite{MWC:SPLC09,LGCR:SPLC15,KHS+:VaMoS24,BSRC10}, even the hardest instances (e.g., for the Linux kernel) can be solved in a few milliseconds.\footnote{after an initial bootstrapping phase} Other analyses and applications, such as model counting~\cite{SHN+:EMSE23,KKS+:ASE22}, sampling~\cite{HSO+:VaMoS24,HFG+:EMSE22,KTS+:VaMoS20}, or knowledge compilation in general~\cite{DM:JAIR02,HSS+:SPLC24,T:SPLC20,DHK:SPLC24,S:SPLC08,SRH+:TOSEM24,SHN+:EMSE23} scale orders of magnitude worse than SAT solving. From the point of view of feature-model analysis, one is, therefore, more interested in preprocessing techniques that preserve the model count or even the configuration space itself~\cite{DHK:SPLC24,LM:AAAI14,KKS+:ASE22}. 

The common objective of such preprocessing techniques is to reduce the number of variables and clauses in a model's CNF representation, while preserving equivalence. For example, approaches that propagated core and dead variables~\cite{HSS+:SPLC24,BSRC10}\footnotetext[3]{In SAT solving, the terms unit and failed variables are used~\cite{BJK21}.}, extracted XOR dependencies~\cite{HSS+:SPLC24,FBAH20}, or applied vivification~\cite{PHS:ECAI08,BJK21,DHK:SPLC24} have been used previously.

Atomic sets are sets of variables that attain the same truth value in all valid configurations~\cite{SKT+:ICSE16,S:SPLC08}. Consequentially, each atomic set can be condensed into a single variable, while preserving equivalence~\cite{S:SPLC08,BSRC10}. However, previous approaches for computing atomic sets can be expensive to compute~\cite{MTS+17,SKT+:ICSE16}, require model counting~\cite{SKH+:AMAI23,SNB+:VaMoS21} or do not consider cross-tree constraints~\cite{BSRC10,S:SPLC08,GHF+:SPLC23,ZZM04}. 

In this work, we present a novel, counting-free algorithm for computing atomic sets based solely on SAT solving. Thereby, we answer the call by \citeauthor{DBS+:SoSyM15}~\cite{DBS+:SoSyM15} for a more efficient algorithm for atomic-set computation. Our evaluation demonstrates that our algorithm outperforms previous approaches by at least an order of magnitude on average. Most notable, our algorithm scales to the infamous Linux feature model, computing the atomic sets of Linux 2.6.33.3 in less than 5\,s and even succeeds for Linux 6.4. Additionally, we demonstrate the utility of computing atomic sets by exploiting them for preprocessing and comparing the results against the state-of-the-art preprocessor \texttt{pmc}~\cite{LM:AAAI14,DHK:SPLC24}.

\begin{figure}
	\vspace{10pt}
	
	\begin{forest}
		for tree = {feature, parent anchor = south, child anchor = north, edge = {plain}, l sep-=1pt,l=0}
		[A, name = A
		[B, name = B, edge = mandatory
		[D, name = D]
		[E, name = E]
		]
		[C, name = C, edge = optional]
		]
		\drawXOR{D}{B}{E};
		\node[below = 52pt of A, font = \sffamily\small] {C $\Leftrightarrow$ E};
		\node[above = 1.5pt of B.north, anchor = east, xshift = 0pt, font = \footnotesize, blue] {\textit{mandatory} $\rightarrow$};
		\node[above = 1.5pt of C.north, anchor = west, xshift = 0pt, font = \footnotesize, blue] {$\leftarrow$ \textit{optional}};
		\node[below = 1pt of B.south east, anchor = north west, xshift = 0pt, font = \footnotesize, blue] {$\leftarrow$ \textit{alternative} group};
	\end{forest}
	
	\caption{Example of a Feature Model}
	\label{fig:fmre}
	
	\vspace*{-10pt}
\end{figure}
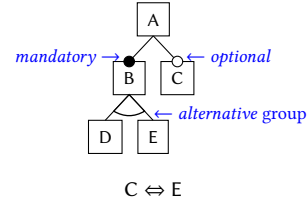

\section{Background}\label{sec:bg}

In this section, we give a brief overview on feature models, their analysis, and give a definition for atomic sets.

\paragraph{Feature Models}

Feature models encode the configuration space of configurable systems in a feature tree and cross-tree constraints~\cite{B05}. Configurations can be derived from a feature model, by selecting and deselecting features. A configuration is valid, when it satisfies all constraints imposed by the model~\cite{BSRC10}.

Consider \autoref{fig:fmre}, which depicts a minimalist feature model consisting of five features \textsf{A}, \textsf{B}, \textsf{C}, \textsf{D}, and \textsf{E}. Selecting a child feature always mandates the selection of its parent feature. Conversely, \textsf{B} is a mandatory child of \textsf{A}, meaning that it must always be selected when \textsf{A} is selected. \textsf{D} and \textsf{E} form an alternative group. Whenever their parent \textsf{B} is selected, exactly one of them must be selected as well. Lastly, \textsf{C} is optional and may be selected freely. However, due to the cross-tree constraint \textsf{C} $\Leftrightarrow$ \textsf{E}, a selection of \textsf{C} will always imply the selection of \textsf{E} and vice versa. We refer to the work of \citeauthor{BSRC10}~\cite{BSRC10} for a more complete introduction to feature models.

\paragraph{Analyzing Feature Models}
In order to use off-the-shelf satisfiability (SAT) solvers, model counters, or other tools, feature models are typically translated into Boolean formulas in conjunctive normal form (CNF)~\cite{BSRC10,SHN+:EMSE23}. In the following, we will denote the set of variables of such a formula $F$ as $V(F)$. A typical analysis is the computation of variables that are \emph{core} and are selected in every valid configuration~\cite{BSRC10}. From the perspective of Boolean formulas, a variable $v$ is core when the conjunction of its negative with $F$ is unsatisfiable, i.e., $\text{UNSAT}(F \land \overline{v})$. Conversely, a variable $v$ is called \emph{dead} when is not selected in any valid configuration, i.e., $\text{UNSAT}(F \land v)$ holds. We denote the sets of these variables as $V_\textit{core}(F), V_\textit{dead}(F) \subset V(F)$, respectively.\footnotemark

\paragraph{Preprocessing}
Preprocessing is a collective term for a multitude of techniques with the common objective of reducing the cost for subsequent computations~\cite{BJK21,KKS+:ASE22,BSRC10}. Therefore, we limit ourselves to a brief discussion of the techniques relevant to our use case: reducing the number of variables and clauses in a model's CNF representation, while preserving equivalence. Central to our approach is \emph{unit propagation}~\cite{BJK21}, where constant variables (i.e., core or dead variables), are replaced by their respective values in all clauses. Subsequently, satisfied clauses and tautologies can be removed from the set of clauses.

\paragraph{Atomic Sets}
A non-trivial set of variables is called \emph{atomic set} if all members attain the same value in every valid configuration~\cite{SKT+:ICSE16}. More formally, a set $A \subset V(F)$ with $\lvert A\rvert \geq 2$ is an atomic set if either $A \equiv V_\textit{core}(F)$, $ A \equiv V_\textit{dead}(F)$, or when it holds that UNSAT($F \land x \land \overline{y}$) and UNSAT($F \land \overline{x} \land y$) for all $x,y \in A$. Note that our example feature model in \autoref{fig:fmre} possesses two atomic sets, namely $\{\textsf{A}, \textsf{B}\}$, as \textsf{B} is a mandatory child of \textsf{A}, and $\{\textsf{C}, \textsf{E}\}$, due to the bi-implication in the cross-tree constraint.

As all variables in an atomic set imply each other, they can be condensed into a single variable $v$~\cite{S:SPLC08,BSRC10}. Consequently, all literal occurrences pertaining to other variables in the atomic set may be replaced by the respective literals of $v$. Afterwards, unit propagation can be used, with satisfied and tautological clauses being removed afterwards, as before. In the following, we refer to this process as \emph{atomic-set elimination} (ASE).

\section{The Insight}

Before we present our algorithm, we briefly discuss the central insight it is built upon. Let $F$ be a Boolean formula and $S$ a satisfying variable assignment for $F$. Furthermore, let $S^+, S^- \subset S$, with $S^+$ denoting the set of variables assigned \textit{true} in $S$, and, analogously, let $S^-$ be the set of variables assigned \textit{false}. Together with the definitions for core and dead variables as well as atomic sets (cf. \autoref{sec:bg}), it follows that $V_\textit{core}(F) \subseteq S^+$, \mbox{$V_\textit{dead}(F) \subseteq S^-$}, and that either $A \subseteq S^+$ or $A \subseteq S^-$ holds for all atomic sets $A$ of $F$.

Moreover, let $S_{x = 1}, S_{x = 0}$ denote variable assignments that satisfy $F$ and contain the assignment of true (1) or false (0) to $x$, respectively. Then, if the variable $v$ is member of an atomic set $A$, it holds that \mbox{$A \subseteq S_{v=1}^+ \cap S_{v=0}^-$}, as all other members of $A$ must attain the same truth value as $v$ in order for $A$ to actually be an atomic set. 

In the following, we apply these insights to the SAT certificates given by state-of-the-art SAT solvers~\cite{Knuth15SAT,IMM:SAT18}, in order to reduce the number of candidate variables when computing atomic sets.

\section{The Algorithm}
Our algorithm is depicted in \autoref{alg:gnt}. It takes a Boolean formula $F$, its set of variables $V(F)$ as well as the externally computed sets of core $V_\textit{core}(F)$ and dead variables $V_\textit{dead}(F)$ as inputs and returns the set of atomic sets. Note that it suffices in practice to only supply the formula and its number of variables, as we will discuss at the end of this section.

At the core, our algorithm \texttt{GnT} (see \autoref{alg:gnt}) follows a generate-and-test strategy based on the intuition that if two variables are part of the same atomic set, they must always be either true or false together. Hence, by computing valid variable assignments for both a literal $v$ (\autoref{alg:sat1}) and its negation $\overline{v}$ (\autoref{alg:sat2}), we generate a set of candidate variables (\autoref{alg:cand}) that fulfill this property. Afterwards, it only remains to verify each candidate to always attain the same truth value as $v$, which can be easily verified by testing with a SAT solver for unsatisfiability. As neither $v$ nor any of the candidate variables are core or dead, it suffices to test for UNSAT($F \land v \land \overline{u}$) and \mbox{UNSAT($F \land \overline{v} \land u$)} (cf. \autoref{alg:unsat}). In particular, it suffices to only test $u$ against $v$ and not all variables in $A$, as all $w \in A$ are implied by $v$, due to $A$ being a subset of an atomic set.

\begin{algorithm}[h]
	
	\SetKw{Continue}{continue}
	\SetKw{Break}{break}
	\SetKwBlock{Loop}{Loop}{end}
	\SetKwFor{For}{for}{do}{endfor}
	
	\SetKwData{Uncovered}{uncovered}
	\SetKwData{Config}{config}
	\SetKw{and}{and}
	
	\SetKwIF{If}{ElseIf}{Else}{if}{then}{else if}{else}{endif}
	
	\SetKwRepeat{Do}{do}{while}
	
	\SetAlgoLined
	\KwIn{formula $F$, sets $V(F)$, $V_\textit{core}(F)$, and $V_\textit{dead}(F)$}
	\KwOut{set $\mathcal{A}$ of atomic sets}
	
	\BlankLine
	
	$\mathcal{A}$ $\leftarrow$ $\emptyset$
	
	\text{decided} $\leftarrow$ $V_\textit{core}(F) \cup V_\textit{dead}(F)$ 
	
	\BlankLine
	
	\ForEach{$v$ $\in$ $V(F) \setminus \text{decided}$}
	{
		$S_{v = 1}$ $\leftarrow$ solve SAT($F \land v$)\label{alg:sat1}
		
		$S_{v = 0}$ $\leftarrow$ solve SAT($F \land \overline{v}$)\label{alg:sat2}
		
		\BlankLine
		
		$C$ $\leftarrow$ $\left\{x \in S_{v=1} ~|~ x > 0 \right\}\cap \left\{|x| ~|~ x \in S_{v = 0},~x < 0 \right\}$\label{alg:cand}
		
		\BlankLine
		
		$A$ $\leftarrow$ $\{v\}$
		
		\ForEach{$u$ $\in$ $C$ $\setminus$ \text{decided}}{
			\If{UNSAT($F\land v \land \overline{u}$ ) ~\and~ UNSAT($F \land \overline{v} \land u$)}{\label{alg:unsat}
				$A$ $\leftarrow$ $A \cup \{u\}$
			}
		}\label{alg:after}
		
		\BlankLine
		
		\If{$\lvert A\rvert > 1$}{
			$\mathcal{A}$ $\leftarrow$ $\mathcal{A} \cup \{A\}$
		}
		
		\BlankLine
		
		\text{decided} $\leftarrow$ $\text{decided} \cup A$
	}
	
	\KwRet{$\mathcal{A}$}
	\caption{``Generate and Test'' Algorithm}
	\label{alg:gnt}
\end{algorithm}

\paragraph{Augmentations}
We omitted some optimizations in the depiction of the algorithm in \autoref{alg:gnt} for better readability. As hinted above, it is not necessary to compute the sets of core and dead variables externally and supply them to the algorithm. Instead, the SAT calls in \autoref{alg:sat1} and \autoref{alg:sat2} can be reused. For dead variables, the SAT call in \autoref{alg:sat1} will always return UNSAT, and vice versa the SAT call in \autoref{alg:sat2} for core variables. While one could use the candidate elimination process in Lines 8--12 to compute the sets of core and dead variables, this would require more SAT calls compared to detecting core and dead variables with the calls in \autoref{alg:sat1} and \autoref{alg:sat2}, without additional benefit.

To reduce the number of SAT calls during the candidate elimination, the proofs from SAT calls returning SAT can be used to eliminate candidates. In particular, let $S_{v\overline{u}}$ be a configuration which refutes the test $\text{UNSAT}(F\land v\land \overline{u})$. Then the set \[\{|w| ~|~ w \in S_{v\overline{u}},~w < 0\}\] contains variables which also not always attain the same value as $v$ and, hence, can be eliminated from the candidate set as well. Conversely for the set $S_{\overline{v}u}$ which refutes the test $\text{UNSAT}(F\land\overline{v}\land u)$.

Finally, the set difference between the candidate set $C$ and the resulting atomic set $A$ can be used to speedup the verification of subsequent atomic sets. Let $R = C \setminus A$ be the set of remaining variables that do not partake in an atomic set with $v$ (cf. \autoref{alg:after}). Note that $A$ contains at least $v$ and, therefore, $R$ is a proper subset of $C$. As a consequence of the construction of $C$, any atomic set containing a variable $u \in R$ must be a subset of $R$ as well, as any variable $w \in V(F) \setminus C$ either had a different truth value to $u$ in $S_{v = 1}$ or $S_{v = 0}$. This knowledge can be used to further reduce the candidate set $C$ before the elimination process, by intersecting $C$ with the respective $R$ after \autoref{alg:cand}.

\section{Evaluation}
In this evaluation, we verify our algorithm and compare its performance against existing implementations in ddnnife~\cite{SRH+:TOSEM24,SLT:ASE24} and FeatureIDE/FeatJar~\cite{MTS+17} on a variety of well-know industrial feature models~\cite{SBK+:SPLC24}. Note that we excluded FlamaPy~\cite{GHF+:SPLC23} as it only considers atomic sets in the feature-model hierarchy and does not consider cross-tree constraints.

In particular, we answer the following research questions:

\resq{1}{(\textit{Correctness}) Does our algorithm correctly compute atomic sets?}
\resq{2}{(\textit{Performance}) How does the performance of our algorithms compare against previous approaches?}
\resq{3}{(\textit{Utility}) Is it worth to eliminate atomic sets for preprocessing?}

\subsection{Preliminaries}

\paragraph{Environment}
The experiment was conducted on a machine with an AMD Ryzen\texttrademark{} 5 8645HS and 16\,GB RAM. Everything was executed under Arch Linux (\texttt{6.12.9-arch1-1}) in a single thread and Python 3.13. 

\paragraph{Model Instances}
We chose 14 well-known real-world feature models (cf. \autoref{tbl:performance}) that are commonly used for benchmarking~\cite{SBK+:SPLC24,HFG+:EMSE22,HSS+:SPLC24,DHK:SPLC24,SHN+:EMSE23}. In particular, we chose four CDL feature models~\cite{LST+:OSR06,BSL+:TSE13} (\textsf{am31\_sim}, \textsf{ea2468}, \textsf{ecos-icse11}, \textsf{p2106}), two automotive models~\cite{KTS+:ESECFSE17} (\textsf{automotive01}, \textsf{automotive02\_v4}), and one financial model~\cite{FAR:SPLC20} (\textsf{financialservices01}). The remaining seven models were extracted from the kconfig modeling language~\cite{BSL+:TSE13, LSB+:SPLC10}.\footnote{The model for Linux 6.4 was extracted with \texttt{torte} (\url{https://github.com/ekuiter/torte})} All instances are available as part of our collection of benchmark instances~\cite{SBK+:SPLC24}.

\paragraph{Tools}
\texttt{ddnnife}\footnote{\url{https://github.com/SoftVarE-Group/d-dnnf-reasoner}} is a d-DNNF reasoner implemented in Rust~\cite{SRH+:TOSEM24,SLT:ASE24}. It compiles a CNF into a d-DNNF and subsequently computes atomic sets using model counting and feature cardinalities.

\texttt{FeatJar}\footnote{\url{https://github.com/FeatureIDE/FeatJAR}} is ``a collection of Java libraries for feature-oriented software development with the goal of eventually replacing FeatureID''~\cite{MTS+17}. It computes atomic sets by mutating SAT solutions and uses Sat4J~\cite{LP:JSAT10} as SAT solver. 

\texttt{pmc}\footnote{\url{http://www.cril.univ-artois.fr/kc/pmc.html}} is a state-of-the-art preprocessor~\cite{LM:AAAI14,DHK:SPLC24} for Boolean formulas in CNF. We use its equivalence-preserving setting.

Our algorithm is implemented in Python as part of the \texttt{ddueruem} project~\cite{HMS+:SPLC22}, which is also used to interface with the other tools via Python's \texttt{subprocess}\footnote{\url{https://docs.python.org/3/library/subprocess.html}} module. We employ PySAT's default SAT solver \texttt{MiniSat} 2.2 as SAT solver.\footnote{\url{http://minisat.se/MiniSat.html}}

\paragraph{Experiment Setup}
For all model instances we attempt to compute the atomic sets within a time limit of 600\,s per model and approach. For \texttt{ddnnife}, we discriminate between the knowledge-compilation time and the time for atomic-set computation. Afterwards, we perform atomic-set elimination (ASE, cf. \autoref{sec:bg}). We record the number of variables and the number of clauses after preprocessing. Finally, we compare the achieved reductions against the reductions achieved by the preprocessor \texttt{pmc}.

\paragraph{Verification}
To verify the correctness of our \texttt{GnT} algorithm, we compare its computed atomic sets against the atomic sets computed by other approaches, whenever the scale. We verify the preprocessing by comparing the model counts and feature cardinalities~\cite{SNB+:VaMoS21} of the preprocessed models against those of the original input models.

\subsection{Results}

\begin{figure}
	\includegraphics[width=\linewidth]{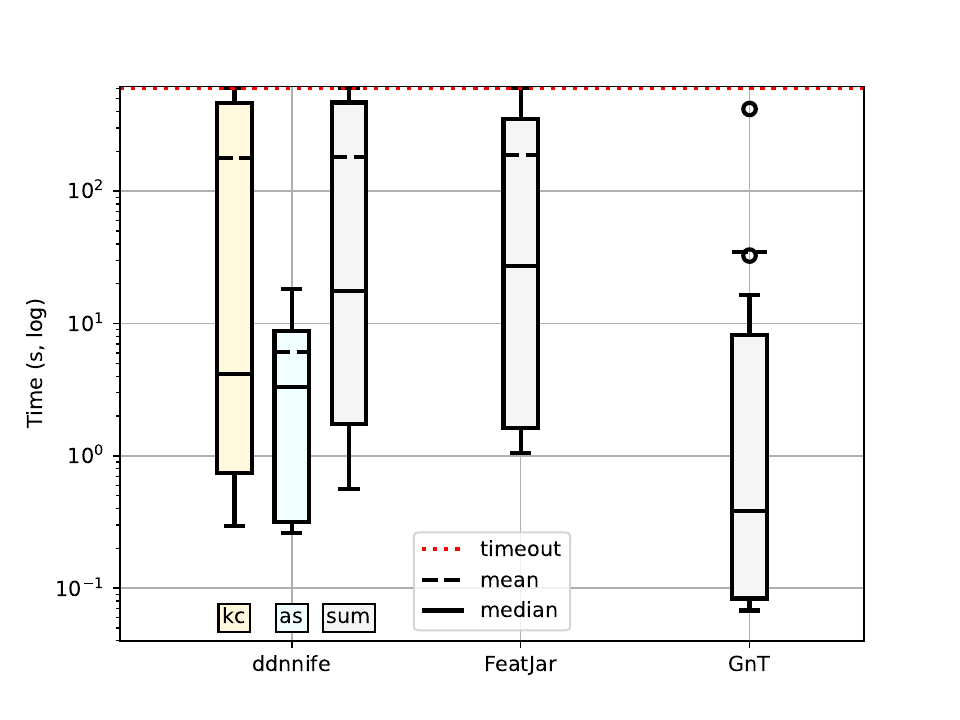}
	\caption{Boxplot of Tool Performances}
	\label{fig:bpperf}
\end{figure}

\begin{table*}
	\caption{Characteristics and Tool Performances for the Evaluated Models}
	\label{tbl:performance}
	\renewcommand{\arraystretch}{1.25}
\smaller
\begin{tabular}{|l|rr|rrrr|rrrrr|}
	\hline
	& & & \multicolumn{4}{c|}{\thead{Atomic Sets}} & \multicolumn{5}{c|}{\thead{Runtimes}} \\
	\multicolumn{1}{|c|}{ \multirow{-2}{*}{ \thead{Instance}} } & \multicolumn{1}{c}{\multirow{-2}{*}{ \thead{\texttt{\#}Variables}}} & \multicolumn{1}{c|}{\multirow{-2}{*}{ \thead{\texttt{\#}Clauses}}} & \thead{\texttt{\#}Sets} & \thead{\texttt{\#}Variables} & \thead{mean} & \thead{max} & \thead{\texttt{ddnnife}} & \thead{(kc)} & \thead{as} & \thead{\texttt{FeatJar}} & \thead{\texttt{GnT}}\\
	\toprule\hline
	\textsf{ embtoolkit } & 1,179 & 5,414 & 27 & 618 & 22.89 & 236 &  0.56  & (0.29)  &  0.27  & 2.07  &  \cellcolor{green!25}   0.07 \\
	\textsf{ busybox\_1.18.0 } & 854 & 1,163 & 21 & 83 & 3.95 & 23 &  0.57  & (0.31)  &  0.26  & 1.06  &  \cellcolor{green!25}   0.08 \\
	\textsf{ financialservices01 } & 771 & 7,238 & 58 & 184 & 3.17 & 22 &  0.72  & (0.42)  &  0.29  & 19.49  &  \cellcolor{green!25}   0.14 \\
	\textsf{ automotive01 } & 2,513 & 10,300 & 167 & 1,008 & 6.04 & 195 &  0.92  & (0.53)  &  0.39  & 34.68  &  \cellcolor{green!25}   0.63 \\
	\textsf{ am31\_sim } & 1,178 & 2,344 & 135 & 724 & 5.36 & 64 &  4.23  & (1.37)  &  2.86  & 1.34  &  \cellcolor{green!25}   0.07 \\
	\textsf{ automotive02\_v4 } & 18,616 & 350,119 & 165 & 2,287 & 13.86 & 1,777 &  \cellcolor{green!25}  7.71  &  (3.91)  &  3.80  &  $\blacklozenge$  &   32.64 \\
	\textsf{ ea2468 } & 1,408 & 2,808 & 155 & 911 & 5.88 & 126 &  13.16  & (4.09)  &  9.07  & 1.90  &  \cellcolor{green!25}   0.09 \\
	\textsf{ ecos-icse11 } & 1,244 & 3,146 & 150 & 765 & 5.10 & 64 &  22.33  & (4.82)  &  17.51  & 1.54  &  \cellcolor{green!25}   0.10 \\
	\textsf{ p2106 } & 1,262 & 2,528 & 141 & 783 & 5.55 & 64 &  22.44  & (4.20)  &  18.24  & 1.50  &  \cellcolor{green!25}   0.08 \\
	\textsf{ embtoolkit-smarch } & 23,516 & 180,511 & 1,456 & 23,221 & 15.95 & 6,561 &  74.66  & (66.81)  &  7.86  & 397.72  &  \cellcolor{green!25}   9.11 \\
	\textsf{ buildroot } & 14,910 & 45,603 & 2,011 & 14,410 & 7.17 & 6,895 &  & ($\blacklozenge$) &   & 169.10  &  \cellcolor{green!25}   5.25 \\
	\textsf{ freetz } & 31,012 & 102,705 & 3,749 & 30,868 & 8.23 & 14,445 &  & ($\blacklozenge$) &   &  $\blacklozenge$  &  \cellcolor{green!25}   16.37 \\
	\textsf{ linux\_2.6.33.3 } & 6,467 & 40,121 & 251 & 1,072 & 4.27 & 310 &  & ($\blacklozenge$) &   & 206.55  &  \cellcolor{green!25}   4.21 \\
	\textsf{ linux\_6.4 } & 47,122 & 281,253 & 1,755 & 23,505 & 13.39 & 16,550 &  & ($\blacklozenge$) &   &  $\blacklozenge$  &  \cellcolor{green!25}   418.54 \\
	
	\hline
\end{tabular} 	

	{\footnotesize times in s, \tikz[baseline=-\the\dimexpr\fontdimen22\textfont2\relax ]{\node[draw = gray, fill = green!25] {};} = best, $\blacklozenge$ = timeout, (kc) = time for knowledge compilation to d-DNNF, as = time for atomic-set computation}
\end{table*}

\autoref{tbl:performance} contains information on the input models and their number of atomic sets, together with the total, mean, and maximum number of variables therein. In addition, the table also reports the runtimes for all approaches, namely \texttt{ddnnife}, \texttt{FeatJar}, and our novel algorithm \texttt{GnT}. Notable, \texttt{GnT} is the only approach that scales to all instances. It outperforms both \texttt{ddnnife} and \texttt{FeatJar} on all instances but \textsf{automotive02\_v4}, for which \texttt{ddnnife} performs significantly better ($\sim$ 5x). All approaches computed identical atomic sets, whenever they scaled.

As expected~\cite{SHN+:EMSE23}, \texttt{ddnnife} times out during knowledge compilation for \textsf{buildroot}, \textsf{freetz}, and both Linux models. When considering the combined runtime of knowledge compilation and atomic-set computation, \texttt{ddnnife} is 74.7x slower on average, with 1.47x (\textsf{automotive01}) being the lowest factor besides \textsf{automotive02\_v4} and 288.29x (\textsf{ecos-icse11}) being the biggest factor. Without the time effort for knowledge compilation, \texttt{ddnnife} is still 56.1x slower on average, but faster for \textsf{automotive02\_v4} (0.12x), \textsf{automotive01} (0.63x), and \textsf{embtoolkit-smarch} (0.86x).

Consider \autoref{fig:bpperf} for further illustration, where we break down the runtime of \texttt{ddnnife} into the knowledge-compilation phase (kc) and the atomic-set computation (as). Even though the boxplot for \texttt{GnT} also contains runtimes for instances to which \texttt{ddnnife} did not scale, one can observe that \texttt{GnT} is an order of magnitude faster than \texttt{ddnnife} on median. This is surprising, as atomic sets can be computed in polynomial time on d-DNNFs~\cite{SNB+:VaMoS21}.

In comparison to \texttt{FeatJar}, which also is a counting-free algorithm based on SAT solving, our algorithm is strictly faster, by a factor of 39.6x on average. Part of this difference may be due to \texttt{FeatJar}'s implementation in Java and their use of Sat4J for SAT solving. As our algorithm does, \texttt{FeatJar} also struggles for instances with many variables, such as the aforementioned \textsf{automotive02\_v4} model. However, it also outperforms \texttt{ddnnife} by a factor of 6.9x on average and also in terms of solved instances, as \texttt{FeatJar} also scales to \textsf{buildroot} and \textsf{linux\_2.6.33.3}.

Based on the numbers and sizes of detected atomic sets, one would expect that ASE has a decisive impact on most models. Indeed, as depicted in \autoref{tbl:preprocessing}, ASE, on average, reduces the number of variables by 46.1\,\% (median: 50.0\,\%) and the number of clauses by 52.7\,\% (median: 56.5\,\%). A notable exception is \textsf{automotive02\_v4} on which ASE only has marginal effect. This, however, does not come as a surprise, as \textsf{automotive02\_v4} is well known to be dominated by alternative groups~\cite{FBAH20,HSS+:SPLC24}. On the other side of the spectrum, the effect of ASE on \textsf{embtoolkit-smarch} and \textsf{freetz}, both hard benchmark models~\cite{HSO+:VaMoS24,DHK:SPLC24}, is astounding. We successfully verified our preprocessing by comparing the model counts and feature cardinalities of original and preprocessed CNFs, where possible~\cite{SHN+:EMSE23}.

Lastly, the comparison with the preprocessor \texttt{pmc} bears astonishing results. For one, ASE outperforms \texttt{pmc} on all but four instances, with \texttt{pmc} performing better for \textsf{automotive01}, \textsf{financialservices01}, and both Linux models. Curiously, both preprocessing approaches appear to be complementary, as combining both approaches achieves the best reductions in the number of clauses. Note that \texttt{pmc} is not necessarily deterministic, hence we ran the deterministic ASE first.

\subsection{Discussion}

\begin{table*}
	\caption{Effect of Preprocessing on the Models' CNFs}
	\label{tbl:preprocessing}
	\renewcommand{\arraystretch}{1.25}
\smaller
\begin{tabular}{|l|rcr|rcr|rr|rr|r|}
	\hline
	\thead{Instance} & \multicolumn{3}{c|}{\thead{\texttt{\#}Variables}} & \multicolumn{3}{c|}{\thead{\texttt{\#}Clauses (ASE)}} & \multicolumn{2}{c|}{\thead{\texttt{\#}Clauses (\texttt{pmc}) $^\blacklozenge$}} & \multicolumn{3}{c|}{\thead{\texttt{\#}Clauses (ASE + \texttt{pmc}) $^\blacksquare$ $^\spadesuit$}}\\
	\toprule\hline
	\textsf{ embtoolkit } & 1,179 & $\xrightarrow{\raisebox{-1ex}{ -50.4\% } }$ & 585 & 5,414 & $\xrightarrow{\raisebox{-1ex}{ -38.3\% } }$ &  \cellcolor{green!25}  3,341 &  3,453 & 3.4\% & 2,768 & -17.2\% & -48.9\%\\
	\textsf{ busybox\_1.18.0 } & 854 & $\xrightarrow{\raisebox{-1ex}{ \phantom{0}-7.5\% } }$ & 790 & 1,163 & $\xrightarrow{\raisebox{-1ex}{ -53.4\% } }$ &  \cellcolor{green!25}  542 &  610 & 12.5\% & 525 & -3.1\% & -54.9\%\\
	\textsf{ financialservices01 } & 771 & $\xrightarrow{\raisebox{-1ex}{ -16.3\% } }$ & 645 & 7,238 & $\xrightarrow{\raisebox{-1ex}{ \phantom{0}-6.9\% } }$ &  6,736 &  \cellcolor{green!25}  5,076 & -24.6\% & 4,605 & -9.3\% & -36.4\%\\
	\textsf{ automotive01 } & 2,513 & $\xrightarrow{\raisebox{-1ex}{ -33.5\% } }$ & 1,672 & 10,300 & $\xrightarrow{\raisebox{-1ex}{ -36.2\% } }$ &  6,573 &  \cellcolor{green!25}  6,094 & -7.3\% & 4,675 & -23.3\% & -54.6\%\\
	\textsf{ am31\_sim } & 1,178 & $\xrightarrow{\raisebox{-1ex}{ -50.0\% } }$ & 589 & 2,344 & $\xrightarrow{\raisebox{-1ex}{ -56.4\% } }$ &  \cellcolor{green!25}  1,022 &  1,987 & 94.4\% & 878 & -14.1\% & -62.5\%\\
	\textsf{ automotive02\_v4 } & 18,616 & $\xrightarrow{\raisebox{-1ex}{ -11.4\% } }$ & 16,493 & 350,119 & $\xrightarrow{\raisebox{-1ex}{ \phantom{0}-5.6\% } }$ &  \cellcolor{green!25}  330,637 &  332,027 & 0.4\% & 329,589 & -0.3\% & -5.9\%\\
	\textsf{ ea2468 } & 1,408 & $\xrightarrow{\raisebox{-1ex}{ -53.7\% } }$ & 652 & 2,808 & $\xrightarrow{\raisebox{-1ex}{ -59.3\% } }$ &  \cellcolor{green!25}  1,142 &  2,340 & 104.9\% & 974 & -14.7\% & -65.3\%\\
	\textsf{ ecos-icse11 } & 1,244 & $\xrightarrow{\raisebox{-1ex}{ -49.4\% } }$ & 629 & 3,146 & $\xrightarrow{\raisebox{-1ex}{ -63.2\% } }$ &  \cellcolor{green!25}  1,159 &  2,119 & 82.8\% & 939 & -19.0\% & -70.2\%\\
	\textsf{ p2106 } & 1,262 & $\xrightarrow{\raisebox{-1ex}{ -50.9\% } }$ & 620 & 2,528 & $\xrightarrow{\raisebox{-1ex}{ -57.4\% } }$ &  \cellcolor{green!25}  1,077 &  2,130 & 97.8\% & 925 & -14.1\% & -63.4\%\\
	\textsf{ embtoolkit-smarch } & 23,516 & $\xrightarrow{\raisebox{-1ex}{ -92.6\% } }$ & 1,750 & 180,511 & $\xrightarrow{\raisebox{-1ex}{ -96.2\% } }$ &  \cellcolor{green!25}  6,800 &  38,326 & 463.6\% & 3,308 & -51.4\% & -98.2\%\\
	\textsf{ buildroot } & 14,910 & $\xrightarrow{\raisebox{-1ex}{ -83.2\% } }$ & 2,511 & 45,603 & $\xrightarrow{\raisebox{-1ex}{ -81.3\% } }$ &  \cellcolor{green!25}  8,549 &  22,526 & 163.5\% & 4,181 & -51.1\% & -90.8\%\\
	\textsf{ freetz } & 31,012 & $\xrightarrow{\raisebox{-1ex}{ -87.4\% } }$ & 3,893 & 102,705 & $\xrightarrow{\raisebox{-1ex}{ -82.3\% } }$ &  \cellcolor{green!25}  18,139 &  44,950 & 147.8\% & 5,563 & -69.3\% & -94.6\%\\
	\textsf{ linux\_2.6.33.3 } & 6,467 & $\xrightarrow{\raisebox{-1ex}{ -12.7\% } }$ & 5,646 & 40,121 & $\xrightarrow{\raisebox{-1ex}{ -49.1\% } }$ &  20,423 &  \cellcolor{green!25}  9,182 & -55.0\% & 7,992 & -13.0\% & -80.1\%\\
	\textsf{ linux\_6.4 } & 47,122 & $\xrightarrow{\raisebox{-1ex}{ -46.2\% } }$ & 25,372 & 281,253 & $\xrightarrow{\raisebox{-1ex}{ -52.1\% } }$ &  134,824 &  \cellcolor{green!25}  91,153 & -32.4\% & 65,589 & -28.0\% & -76.7\%\\
	
	\hline
\end{tabular} 	

	{\footnotesize $\blacklozenge$ = compared to ASE, $\blacksquare$ = compared to best of ASE, \texttt{pmc}, $\spadesuit$ = total reduction}
\end{table*}

The answer to \textbf{RQ1} ``Does our algorithm correctly compute atomic sets?'' is straightforward. For each model, the atomic sets computed by the different tools were identical, whenever the respective tools scaled. The additional verification of the individual atomic sets with a SAT solver also succeeded.

With regard to \textbf{RQ2} ``How does the performance of our algorithms compare against previous approaches?'', we conclude that our algorithm compares very favorably. Our \texttt{GnT} algorithm clearly outperforms both \texttt{ddnnife} and \texttt{FeatJar} in both the number of handled instances and performance per instance. In fact, \texttt{GnT} is only outperformed on one model (\textsf{automotive02\_v4}) by \texttt{ddnnife}, but outperforms the others by more than one order of magnitude on average.

Most notable is our answer to \textbf{RQ3} ``Is it worth to eliminate atomic sets for preprocessing?''. For the majority of instances, we were able to compute their atomic sets within 10\,s, a negligible effort compared to the decisive impact ASE has on the CNFs and the expected costs of subsequent tasks, such as knowledge compilation to d-DNNF~\cite{DM:JAIR02,SHN+:EMSE23,SRH+:TOSEM24} or binary decision diagrams~\cite{S:SPLC08,HSS+:SPLC24}. Not only did ASE drastically reduce the number of variables and clauses for the majority of instances, it also outperformed the state-of-the-art preprocessor \texttt{pmc} in this regard. We conclude that ASE appears to be a promising equivalence-preserving preprocessing technique.

\subsection{Threats to Validity}

In the following, we discuss threats to the internal and external validity of our work~\cite{WRH+12}.

\paragraph{Internal Validity}

Our time measurements may be distorted by a number of factors, such as non-determinism, solver warm-up, or thermal throttling of the system. However, our approach is deterministic and a repetition of the evaluation on a second system only showed negligible deviations.

The atomic-set computation or the preprocessing may be erroneous. While we would argue that correctness-by-construction arguments could be made for both the atomic-set computation and the preprocessing, we nevertheless verified their respective outcomes as discussed above. All results were successfully verified.

\paragraph{External Validity}

The performance of the evaluated tools and the impact of preprocessing may not necessarily translate to other feature models. Nevertheless, we would argue that our choice of feature models allows for representative insights. All models in our evaluation are commonly used for benchmarking~\cite{HSS+:SPLC24,DHK:SPLC24,HFG+:EMSE22,FBAH20,SBK+:SPLC24}, stem from a variety of domains and origins~\cite{SBK+:SPLC24}, and possess a variety of properties. We even included a recent model of the Linux kernel (\texttt{linux\_6.4}), to explore the limits of our approach. While we could additionally verify our claims with statistical tests, we refrained from doing so, as the major claims are supported by at least an order of magnitude and the outcome of our experiment is one-sided.

\section{Related Work}

Even though the exploitation of atomic sets for means of preprocessing was proposed two decades ago~\cite{S:SPLC08,ZZM04}, we are - to the best of our knowledge - the first to conduct an empirical analysis on both the computation cost and the preprocessing potential of atomic sets.
Moreover, we are also the first to present a concrete algorithm that also accounts for cross-tree constraints. Previous works by \citeauthor{ZZM04}~\cite{ZZM04} and \citeauthor{S:SPLC08}~\cite{S:SPLC08}, and also the contemporary framework FlamaPy~\cite{GHF+:SPLC23},\footnote{\url{https://github.com/flamapy/fm_metamodel/issues/101}} only account for atomic sets stemming from the feature hierarchy (i.e., from mandatory features). 

Both \citeauthor{DBS+:SoSyM15}~\cite{DBS+:SoSyM15} and \citeauthor{SKT+:ICSE16}~\cite{SKT+:ICSE16} give definitions that account for such atomic sets, but leave the question of efficiently computing such atomic sets open. \citeauthor{SKT+:ICSE16} even claim that computing of atomic sets ``[does] not scale for large feature models''~\cite{SKT+:ICSE16}, which we overcame in this work.

In previous work, we outlined an algorithm for computing atomic sets based on model counting~\cite{SNB+:VaMoS21}, which is implemented by \texttt{ddnnife}. While \citeauthor{SKT+:ICSE16} do not provide an algorithm, they describe how atomic sets may be computed on decomposed feature models~\cite{SKT+:ICSE16}.

\section{Conclusion}

We presented \texttt{GnT}, a novel, counting-free algorithm for computing atomics sets using only SAT solving. Our evaluation shows that \texttt{GnT} outperforms the state of the art, namely \texttt{ddnnife} and \texttt{FeatureIDE}/\texttt{FeatJar} by at least an order of magnitude on average and is the only approach that scales to all models in our evaluation, including \textsf{linux\_2.6.33.3} and \texttt{linux\_6.4}.

Moreover, we demonstrated that atomic-set elimination is a promising preprocessing technique that interleaves well with the existing preprocessing approach of \texttt{pmc}. Thereby, this work builds a foundation for improving the scalability of feature-model analyses, including knowledge compilation, in the future.

\bibliographystyle{ACM-Reference-Format}

\end{document}